\documentclass[prb,twocolumn,aps,showpacs]{revtex4}
\usepackage{graphicx,epsfig}
\usepackage{bm}
\usepackage{amssymb} 
\begin{document}

\title{Hyperfine-induced spin relaxation of a diffusively moving carrier in low dimensions:
implications for spin transport in organic semiconductors}

\author{V. V. Mkhitaryan and V. V. Dobrovitski}

\affiliation{Ames Laboratory, Iowa State University, Ames, Iowa
50011, USA}

\begin{abstract}

The hyperfine coupling between the spin of a charge carrier and
the nuclear spin bath is a predominant channel for the carrier
spin relaxation in many organic semiconductors. We theoretically
investigate the hyperfine-induced spin relaxation of a carrier
performing a random walk on a $d$-dimensional regular lattice, in
a transport regime typical for organic semiconductors. We show
that in $d=1$ and $d=2$ the time dependence of the
space-integrated spin polarization, $P(t)$, is dominated by a
superexponential decay, crossing over to a stretched exponential
tail at long times. The faster decay is attributed to multiple
self-intersections (returns) of the random walk trajectories,
which occur more often in lower dimensions. We also show,
analytically and numerically, that the returns lead to sensitivity
of $P(t)$ to external electric and magnetic fields, and this
sensitivity strongly depends on dimensionality of the system
($d=1$ vs. $d=3$). Furthermore, we investigate in detail the
coordinate dependence of the time-integrated spin polarization,
$\sigma(\mathbf{r})$, which can be probed in the spin transport
experiments with spin-polarized electrodes. We demonstrate that,
while $\sigma(\mathbf{r})$ is essentially exponential, the effect
of multiple self-intersections can be identified in transport
measurements from the strong dependence of the  spin decay length
on the external magnetic and electric fields.
\end{abstract}

\pacs{72.25.Dc, 75.76.+j, 85.75.-d}

\maketitle

\section{Introduction}

Spin dynamics of the charge carriers in organic semiconductors
have attracted much attention recently \cite{ShinarReview,VardShi,
Parmanik07, Drew09, Dediu09, ThoNatMat, Molenkamp11, SVC60,
Molenkamp13, Dediu13, Watanabe14}. On one hand, the processes
which underlie the use of these systems in the organic
light-emitting diodes and organic photovoltaic solar cells, are
explicitly spin-dependent, so that understanding of the spin
dynamics in organic semiconductors is of fundamental interest for
such applications. On the other hand, the long spin lifetimes of
the carriers in the organic semiconductors make them an
interesting candidate for prospective spintronic applications, and
the detailed understanding of the mechanisms of the spin
relaxation is required.

Typically, charge transport in organic semiconductors occurs via
random (incoherent and inelastic) hopping of the polarons carrying
positive or negative charge between localized molecular sites.
During the waiting time between two consecutive hops the carrier
spin interacts via hyperfine coupling with the spins of the nuclei
(mostly, hydrogen), which surround the host site. Thus, the spin
of the carrier  waiting for the next hop undergoes rotation around
a random axis by a random angle at each site. The principal role
of this mechanism was experimentally confirmed \cite{ThoNatMat}
and supported by a number of spin transport \cite{Dediu09,
BobHypInd} and spin resonance \cite{LupBoePRL, ShinarReview,
Malissa} measurements, as well as theoretical works.
\cite{BobSpinDiff,FlattePRL12,RR13,RaikhSpinRel} Another source of
spin relaxation is the spin-orbit interaction, which leads to the
spin-flip scattering in the course of a hop, but this interaction
in organic semiconductors is weak, so the hyperfine-induced spin
relaxation is likely to be the main source of depolarization,
although other theoretical \cite{Yu11} and experimental
\cite{Parmanik07,Watanabe14}  studies favor the spin-orbit
mechanism, and theoretical efforts are made to explain this
controversy. \cite{FlattePRL13}

Spin dynamics of the polarons in low-dimensional systems is of
particular interest: in the polymer-based devices the carriers
mostly move along the 1D polymer chains, while hopping from one
polymer chain to another happens mostly at the intersections.
One-dimensional organic polymer wires can be prepared and their
properties can be studied in detail, see e.g.\
Refs.~\onlinecite{Mahato,Reecht}.
Also, engineering low-dimensional systems is a promising way to
design organic materials with large magnetoresistance,
\cite{BobbertPRB12} which are of much technological interest.

This motivates the theoretical study, described below, of the
hyperfine-induced spin relaxation of a carrier diffusing via
random walk in $d=1$ and $d=2$ dimensions. We investigate in
detail not only the (space-) integrated spin polarization $P(t)$
(which has been addressed in some previous works
\cite{RaikhSpinRel, CzK, LeDoussal, Reineker}), but also the
time-integrated polarization $\sigma({\bf r})$ at the given point
$\bf r$ in space, which can be measured in spin-transport
experiments.
 \cite{VardShi, Parmanik07, Drew09,
Dediu09, ThoNatMat, Molenkamp11, SVC60, Molenkamp13, Dediu13,
Watanabe14, Akselrod} We show that for low-dimensional transport,
these two quantities are related to each other in a rather
non-trivial way. Moreover, we analyze the spin decay length $l_S$
for $d=1$ and $d=2$, and show that it is very sensitive to both
electric and magnetic fields; this is important both for
applications and for the fundamental studies of the transport in
organic semiconductors.

The average polaron hopping rate $\nu$, corresponding to typical
mobilities of $10^{-8}$ to $10^{-6}$ cm$^2$V$^{-1}$s$^{-1}$, is of
the order of 1--100~GHz. \cite{BobMob} At the same time, the
average hyperfine-induced spin precession frequency is of order
$100$~MHz. \cite{ThoNatMat,BobSpinDiff} Therefore, the hopping is
much faster than the hyperfine precession, and the carrier
performs many random-walk steps before its spin polarization
averages out to zero. Thus, the spin relaxation should be
sensitive to the statistics of the underlying random walk. The
particularly important feature of the random walk is the frequency
of returns of the hopping carrier to the same site, i.e. the
frequency of the self-intersections of the random walk
trajectories. When the returns are absent, in the so-called
transient diffusion regime\cite{Feller}, the random hyperfine
field acting on the carrier's spin has no memory: the local
hyperfine environments at different sites are uncorrelated, and
the carrier hops from one site to another never coming back. This
corresponds to the motional narrowing regime of the spin
relaxation \cite{Kuboold} and leads to the exponential decay of
the space-integrated spin polarization,
$P(t)\propto\exp(-t/t_{S,{\rm tr}})$, where $t_{S,{\rm tr}}$ is
the spin relaxation time for the transient-diffusion regime
\cite{Gordon, SchulWol}. The opposite scenario with frequent
returns, known as the persistent diffusion, takes place when the
carriers move in low-dimensional systems. The frequent returns to
the same site lead to the random hyperfine field with long memory
(long-time correlations), which suppresses the motional narrowing
and leads to faster spin relaxation. The role of the memory of the
hyperfine field has been studied for some scenarios, and has been
found to change the spin relaxation of the spin polarization in
$d=1$ case to $P(t)\propto\exp{[-(t/t_{S,1})^{3/2}]}$ at short
times \cite{RaikhSpinRel,CzK,LeDoussal,Reineker}, with the decay
time $t_{S,1}$ depending on the particular model for the hyperfine
fields. Also, it has been noticed that the returns make $P(t)$
very sensitive to the external magnetic field.\cite{RaikhSpinRel}

\begin{figure}[t]
\centerline{\includegraphics[width=75mm,angle=0,clip]{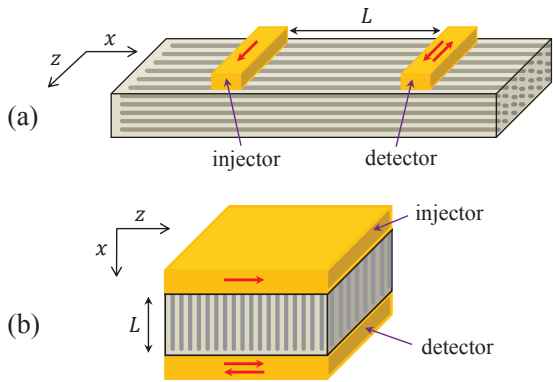}}
\caption{(Color online) Lateral (a) and vertical (b) spin valves
with organic active layer (gray), in which the charge carriers are
restricted to move in one-dimensional channels along the
$x$-direction. The separation between the injecting and detecting
electrodes is $L$.} \label{latvert}
\end{figure}


However, the detailed knowledge about the carrier's spin
relaxation in the case of low-dimensional transport is still
largely lacking, and our work aims at filling this gap. We
investigate analytically and numerically the spin relaxation for
lateral and vertical spin valves (Fig.~\ref{latvert}), with the
carriers moving along the 1D current-carrying channels. We
consider the limit of small current density, when each carrier
moves independently of others, thus working within the
single-particle framework. For the lateral spin valve,
Fig.~\ref{latvert}(a), the spin carrier hops along the very long
linear chain, performing an unbounded random walk. For the
vertical spin valve, Fig.~\ref{latvert}(b), the random walk
happens over finite-size chain with reflecting boundaries. In both
cases we assume that the carrier is injected with the spin state
``up'' at $x=0$, and its spin is probed by the detector at  $x=L$.

We study the space-integrated spin polarization
$P(t)$ and the time-integrated polarization $\sigma({\bf r})$ at
the given point $\bf r$, which can be measured with the detector
lead at a given location. We demonstrate that these quantities
exhibit remarkable universal features, and that
the returns in the course of the carrier diffusion play an
important role in this relation.

First, for the time decay of $P(t)$, we demonstrate the evidence
of the universal scaling: for different values of the hyperfine
coupling strength $b_{\rm hf}$ and hopping rate $\nu$, the decay
of $P(t)$ follows the same curve which depends only on the
normalized dimensionless time $\tau=(\nu t) (b_{\rm
hf}/\nu)^{4/3}$. This scaling holds not only for short times,
where $P(t)$ follows the previously known decay law
$P(t)=\exp{[-(t/t_S)^{3/2}]}$, but also at long times, where we
find a previously unnoticed stretched-exponential decay
$P(t)\sim\exp\bigl(-\alpha_1 t^{3/4}\bigr)$ (for both spin valve
geometries). This scaling holds for both lateral and vertical
spin-valve geometries, at finite magnetic and electric field.
Also, we found that $P(t)$, besides the known sensitivity to the
magnetic field, is also very sensitive to the electric field. We
have observed the similarly strong effect of the returns for the
spin relaxation in $d=2$, leading to the logarithmic corrections
to the standard exponential decay of $P(t)$, and strong
sensitivity to the external fields.

Second, we studied the time-integrated polarization
$\sigma(\mathbf{r})$, which is of much importance for the
spin-dependent transport measurements. We are not aware of any
analytical theory for this quantity, but our numerical studies
reveal unexpected universality in its behavior. For persistent
diffusion in low dimensions, in contrast to the transient
diffusion in 3D, the quantity $\sigma(\mathbf{r})$ is not directly
related to $P(t)$. Our numerical results show that, despite the
essentially non-exponential decay of $P(t)$, the spin transport
decay is exponential, $\sigma(r)\propto \exp(-r/l_S)$, with high
accuracy, even in the presence of the external magnetic and
electric fields. However, the resemblance to the usual 3D
transient-diffusion result is superficial: for both $d=1$ and
$d=2$ cases, the dependence of the spin decay length $l_S$ on the
external fields is very strong, in contrast to the standard 3D
diffusion.

Our results suggest that the character of the carrier diffusion in
an organic semiconductor can be studied in spin transport
experiments, via the field dependence of $l_S$, and vice versa,
the spin transport measurements in low-dimensional organic
semiconductors can be used for accurate sensing of electric and
magnetic fields, and for other similar spintronic applications.

The rest of the paper is organized as follows. In the next Section
we discuss the formulation of the problem and the methods used for
analytical and numerical studies. In Sections \ref{SecLat} and
\ref{SecVert} we consider the spin relaxation for the lateral and
the vertical spin-valve geometries, respectively. Section
\ref{SecTwoD} outlines our results on the spin relaxation in
$d=2$. Details of the analytical calculations are presented in two
Appendices.

\section{Model for the carrier spin relaxation}
\label{secDesc}

\begin{figure}[t]
\centerline{\includegraphics[width=90mm,angle=0,clip]{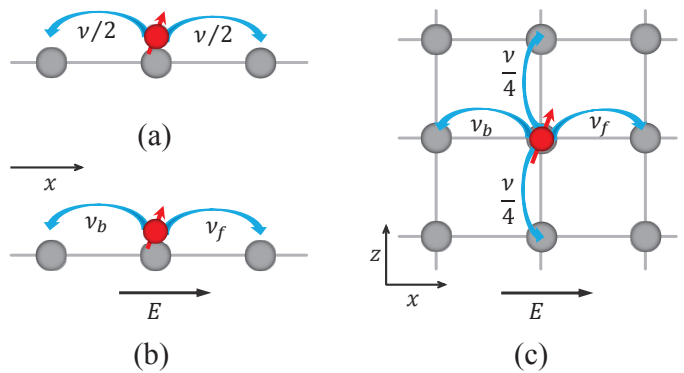}}
\caption{(Color online) Diffusive random walk of a polaron (red
circle) over $d=1$ and $d=2$ dimensional regular lattices of
molecular sites (gray circles).
(a) In the absence of external electric field, polaron hops
between the nearest neighbor sites of $d=1$ linear chain occur
with the equal rates, $\nu/2$, indicated by the blue arrows. (b)
In the presence of external electric field along the linear chain,
hopping rates $\nu_f$ and $\nu_b$, respectively in the directions
forward and backward to the electric field, are different;
$\nu_f>\nu_b$. (c) With an external electric field along the $x$
axis of a regular lattice in $d=2$, hopping rates in the forward
and backward directions to $\hat{\bf x}$, $\nu_f$ and $\nu_b$, can
depend on the electric field, while the rates in the perpendicular
direction are the same as without the field, and equal to $\nu/4$
each.} \label{chain}
\end{figure}

We consider a carrier hopping between sites, see
Fig.~\ref{chain}(a), which model organic molecules or conjugated
segments of polymers. Everywhere below, for both $d=1$ and $d=2$,
we enumerate sites by the integer variable $\bf r$, so that e.g.\
for 1D chain the physical coordinate of the site is $x=ar$, where
$a$ is the distance between the sites.

When the polaron is localized at the site
with the radius-vector $\mathbf{r}$ its spin interacts with $N$
nuclei $\mathbf{I}_{\mathbf{r} k}$ ($k=1,..,N$) surrounding the
given site. Below we assume that all nuclei have spin 1/2, since
the protons in many organic semiconductors are the most abundant
species with the largest nuclear magnetic moment. The Hamiltonian
governing the spin dynamics of the carrier localized at the site
$\bf r$ is:
\begin{equation}\label{locHam}
H_{\mathbf{r}}=B S_z + {\bf S}\,\sum_{k=1}^N
a_{\mathbf{r} k}{\bf I}_{\mathbf{r} k},
\end{equation}
where $a_{\mathbf{r} k}$ is the hyperfine coupling constant between
the carrier spin and nuclear spin $\mathbf{I}_{\mathbf{r} k}$, and
$B$ is the Larmor frequency of the carrier spin
$\mathbf{S} $ in an external magnetic field along the $z$ axis
(everywhere below we take $\hbar=1$ and the
electron's gyromagnetic ratio $\gamma_e=1$, omitting the
difference between the magnetic fields and the Larmor
frequencies).
Theoretical approach to the spin evolution in organic
semiconductors customarily relies on the approximation where the
quantum hyperfine field given by the sum in Eq. (\ref{locHam}),
\begin{equation}\label{Bi}
\hat{\bf b}_{\mathbf{r}}=\sum_{k=1}^Na_{\mathbf{r} k}
\mathbf{I}_{\mathbf{r} k},
\end{equation}
is approximated as a static classical vector ${\bf b}_{\bf r}$ of
random amplitude and direction, sampled from the Gaussian
distribution with zero mean and the standard deviation equal to
$b_{\text{hf}}= \frac12 \sqrt{\sum_ka_{\mathbf{r} k}^2}$; this
approximation is justified by the large number of nuclear spins
coupled to the carrier spin at a given site ($N$ of order $10$ or
more)\cite{SchulWol,ErlNaz,AlHas}. We also assume that the
hyperfine fields at different sites are uncorrelated, so that
$\langle  b^\alpha_\mathbf{r}
b^\beta_{\mathbf{r}'}\rangle_{\text{hf}} = b_{\text{hf}}^2\,
\delta_ {\alpha \beta} \delta_{\mathbf{r} \mathbf{r}'}$, where
$\alpha,\beta=x,y,z$.

When we consider a carrier with the initial spin along
$\hat{\mathbf{z}}$, performing a random walk with trajectory
$\mathbf{r}(t)$, the carrier's spin ${\boldsymbol \mu}(t)$ evolves
according to the equation of motion
\begin{equation}\label{mueq}
\dot{\boldsymbol \mu} = {\bf b}_{\mathbf{r}(t)} \times{\boldsymbol
\mu}(t) = {\hat \Omega}(\mathbf{r}(t)){\boldsymbol \mu}(t)
\end{equation}
where the matrix
\begin{equation}\label{ssm}
{\hat\Omega} (\mathbf{r})\!=\!\!
\left(\!\begin{array}{ccc}
0&- b^z_\mathbf{r}& b^y_\mathbf{r}\\
 b^z_\mathbf{r}&0&- b^x_\mathbf{r}\\
- b^y_\mathbf{r}& b^x_\mathbf{r}&0
\end{array}\!\right)\! ,
\end{equation}
describes the spin rotation taking place at the site $\bf r$.
Formal solution of this equation can
be written in terms of the time-ordered exponent,
$$
{\boldsymbol \mu}(t)=T \exp\!\int_0^t \!dt'{\hat\Omega}
 \bigl( \mathbf{r}(t')\bigr){\boldsymbol \mu}(0),
$$
with the initial condition ${\boldsymbol \mu}(0)=\hat{\mathbf{z}}$.
The spin polarization is obtained by double averaging of the
$z$-component of ${\boldsymbol\mu}(t)$,
\begin{equation}\label{dbav}
P(t)= \langle\langle \mu^z(t)\rangle\rangle\equiv \Bigl \langle
\Bigl\langle T\exp\!\!\int_0^t\!\!dt'\,{\hat\Omega} \bigl(
\mathbf{r}(t')\bigr) \Bigr \rangle _{\text{hf}}\Bigr\rangle
_{\text{rw}}{\Big |}_{zz} \!,
\end{equation}
where $\langle\cdot\rangle_{\text{rw}}$ denotes averaging over the
random walk trajectories, $\langle\cdot\rangle_{\text{hf}}$
denotes averaging over the local hyperfine fields, and the indices
$zz$ denote that we need to take the $zz$ entry of the matrix
which results after the averaging of the time-ordered matrix
exponent.

Without returns, all spin rotations at different sites would be
independent and uncorrelated, leading to exponential decay of the
polarization as a function of time (the motional narrowing
regime). In the presence of the returns, the rotations at
different moments of time are correlated, and the polarization
decay accelerates. In $d=1$, the number of returns of the charge
carrier to a given site after $n$ hops is ${\cal
O}\bigl(n^{1/2}\bigr)$, whereas in $d=2$ and $d=3$ this number is
${\cal O}\bigl( \ln n \bigr)$ and ${\cal O}\bigl( 1\bigr)$,
respectively. \cite{MontWeiss} Therefore one should expect that
the influence of the returns is strong in $d=1$, modest in $d=2$,
and weak in $d=3$ dimensions; below we concentrate mainly on the
$d=1$ case where the effect is strongest.

The exact solution of Eq.~(\ref{dbav}) in the presence of returns
is not available. In order to approach the problem, everywhere
below we employ the fact that the hopping is much faster than
rotation in the hyperfine field, so the parameter
$\eta=b_{\text{hf}}/\nu$ is small. Indeed, the value of
$b_{\text{hf}}$ is typically of order of 100~MHz, while the average carrier hopping rate $\nu$ is about
1--100~GHz, so
$$
\eta=b_{\text{hf}}/\nu \sim 0.1-0.001\ll 1.
$$
We can calculate $P(t)$ via the cumulant expansion in terms of the
small parameter $\eta$:
$$
P(t)= \Bigl\langle \!\Bigl\langle T \exp\!\int_0^t\!dt' \hat{
\Omega}\bigl(\mathbf{r}(t')\bigr)\Bigr \rangle\! \Bigr \rangle
{\Big |}_{zz} \!\!=\exp\!\left(\sum_n K_n(t)\right)\!,
$$
where $K_n(t)$ is proportional to $\eta^n$. The odd
cumulants vanish (since the local Gaussian distributions of ${\bf
b}_\mathbf{r}$ have zero mean), the first non-vanishing cumulant
$K_2(t)$ will determine the behavior of $P(t)$, at least at short
times. We will demonstrate the accuracy of this approach by
comparing the analytically calculated $K_2(t)$ with the results of
the direct numerical simulations.

The numerical simulations are even more important for studying the
time-integrated polarization $\sigma({\bf r})$. We are not aware
of any analytical theory for this quantity which would provide
insights and guide our investigation. Thus, we rely solely on the
numerical results, which demonstrate surprising and interesting
universal features of $\sigma({\bf r})$.

One could do numerical simulations by Monte-Carlo sampling of the
random-walk trajectories and the distributions of the local
fields, thus calculating the average in Eq.~(\ref{dbav}) directly.
However, our results show that the statistical error is quite
large, so instead we employ the approach based on the Liouville
equation.

We describe the carrier spin via its density matrix
$\rho_{\mathbf{r}}(t) = \frac 12\bigl( q_{\mathbf{r}}(t) +
\mathbf{m}_{\mathbf{r}}(t) {\boldsymbol\sigma}\bigr)$, where
$q_{\mathbf{r}}(t)$ is the probability to find the carrier at site
$\mathbf{r}$ at time $t$, and $\mathbf{m}_{\mathbf{r}}(t)$ is its
spin polarization; ${\boldsymbol \sigma}$ is the vector of Pauli
matrices. The carrier dynamics obeys the master equation
\begin{equation}\label{MEp}
\frac {dq_{\mathbf{r}}}{dt} = \sum_{\mathbf{r}'} \bigl[
W_{\mathbf{r}', \mathbf{r}} q_{\mathbf{r}'}(t) -
W_{\mathbf{r},\mathbf{r}'} q_{\mathbf{r}}(t)\bigr],
\end{equation}
where $W_{\mathbf{r}, \mathbf{r}'}$ is the hopping rate from site
$\mathbf{r}$ to $\mathbf{r}'$. At the same time, the spin
polarization follows the generalized drift-diffusion equation,
\begin{equation}\label{MEm}
\frac {d\mathbf{m}_\mathbf{r}}{dt} = \sum_{\mathbf{r}'} \bigl[
W_{\mathbf{r}', \mathbf{r}}\mathbf{m}_{\mathbf{r}'}(t) -
W_{\mathbf{r}, \mathbf{r}'} \mathbf{m}_\mathbf{r}(t)\bigr] + {\bf
b}_\mathbf{r} \times\mathbf{m}_\mathbf{r}.
\end{equation}
For the charge carrier initially injected at the site
$\mathbf{r}=0$ in the spin-up state, the initial conditions
correspond to $q_\mathbf{r}(0)=\delta_{\mathbf{r},0}$ and
$\mathbf{m}_\mathbf{r}(0) = \delta_{\mathbf{r},0} \mathbf{m}(0)$
with $\mathbf{m}(0)=(0,0,1)$ (directed along the $z$-axis).
The solution ${\bf m_r}(t)$ of Eq.~(\ref{MEm}) includes averaging
over the random-walk trajectories of the duration $t$, but the
hyperfine fields at each site are taken as having some specific
directions and amplitudes, i.e.\ the set $\{{\boldsymbol b}_{\bf
r}\}$ of the local fields is fixed. Averaging over the local
hyperfine fields is performed via Monte-Carlo sampling of
$\{{\boldsymbol b}_{\bf r}\}$ at each site from the Gaussian
distribution ${\cal N}({\bf b})= (2\pi
b_{\text{hf}})^{-3/2}\exp\bigl(-|{\bf b}|^2/2b_{\text{hf}}^2
\bigr)$.

In particular, in this way we determine the time-integrated spin
polarization at a given location,
$$
\sigma(\mathbf{r})= \nu\int_0^\infty dt\langle m_\mathbf{r}^z (t)\rangle_{\text{hf}},
$$
which plays an important role in the spin transport measurements.

The effect of the external magnetic field is included into our
model by simply adding the external field $\bf B$ to the local
hyperfine fields. The external electric field $E$ applied along
the $x$-axis (Figs.~\ref{latvert} and \ref{chain}) is taken into
account by modifying the hopping rates: the hops along the field
are more probable than backwards. Below, we assume, in the spirit
of the Miller-Abrahams theory, \cite{M-A} that the backward
hopping rate $\nu_b$ (upwards in the electric field potential) is
exponentially suppressed in comparison with the forward hopping
rate $\nu_f$ (downwards in the electric potential), i.e.\
$\nu_b=(\nu/2) \exp(-\varepsilon)$, where $\varepsilon=eEa/k_BT$
with the Boltzmann constant $k_B$ and temperature $T$, and $eEa$
is the electric potential difference between two neighboring
sites, while the forward-hopping rate remains unchanged,
$\nu_f=\nu/2$. In particular, for $d=1$, we have
\begin{eqnarray}
\label{eqwrr} W_{r, r'}&=& (\nu/2)\,\delta_{r,r'-1} +
(\nu/2)\,\delta_{r,r'+1} \quad {\rm for\ } E=0,\\ \nonumber W_{r,
r'}&=& (\nu/2)\,\delta_{r,r'-1} + (\nu
e^{-\varepsilon}/2)\,\delta_{r,r'+1} \quad {\rm for\ } E\ne0.
\end{eqnarray}

\section{Spin relaxation in a lateral spin valve}
\label{SecLat}

We neglect the effect of injector/detector electrodes, assuming
insignificant tunneling between the leads and the semiconductor.
For the lateral spin valve this implies unbounded diffusion over
an infinite chain. For numerical simulations, we used a long chain
with periodic or reflecting boundary conditions; the length was
large enough to ensure vanishing population near the ends at all
times. We also excluded from consideration the additional
spin relaxation which is possible at the interface between a
ferromagnetic electrode and an organic active layer \cite{Dediu02,
Cinchetti09, Barraud10}.

\subsection{$P(t)$ and $\sigma(r)$ in the absence of external fields}

\begin{figure}[t]
\vspace{-0.4cm}
\centerline{\includegraphics[width=90mm,angle=0,clip]{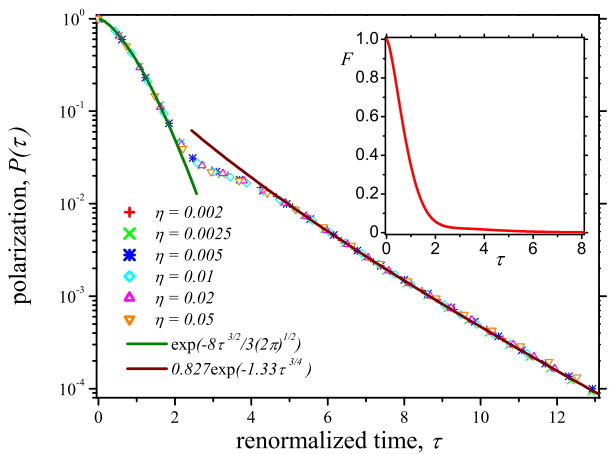}}
\caption{(Color online) Logarithmic plot of $P(t)$ in $d=1$, for
values of $\eta\equiv b_{\text{hf}}/\nu$ ranging from $0.002$ to
$0.05$. When plotted against the renormalized time, $\tau= (\nu
t)\eta^{4/3}$, all data points fall on a single scaling curve. The
resulting scaling curve is well fitted by $\exp
(-8\tau^{3/2}/3\sqrt{2\pi})$ for small $\tau$ and $0.827
\exp(-1.33\tau^{3/4})$ for larger $\tau$. Normal plot of the
scaling curve, $F(\tau)$, is shown in the inset.} \label{scaling}
\end{figure}

Important insights about the short-time behavior of $P(t)$ can be
obtained analytically, using the lowest orders of the cumulant
expansion in terms of the small parameter
$\eta=b_{\text{hf}}/\nu$,
$$
P(t)=\exp\left(\sum_n K_n(t)\right)\approx \exp
{\bigl[K_2(t)\bigr]}.
$$
For a random walk over an infinite chain the first two
non-vanishing cumulants, $K_2$ and $K_4$, are calculated in
Appendix \ref{AppB}. The second cumulant $K_2$ is determined by
the two-time correlation function, $\langle \langle {\hat\Omega}
({\bf r}(t_1))\, {\hat\Omega} ({\bf r}(t_2))\rangle \rangle$. From
the large-$t$ asymptotics of this correlation function we find:
\begin{equation}\label{seccum}
K_2(t)\simeq-\frac{2\,b_{\text{hf}}^2} {\sqrt{2\pi\nu}}
\int\limits_0^t\!dt_1\! \int\limits_0^{t_1}\!dt_2 \frac1
{\sqrt{t_1-t_2}}= -\frac {8\,\eta^2 (\nu t)^{\frac32}}
{3\sqrt{2\pi}} .
\end{equation}
The next non-vanishing cumulant, $K_4$, is determined by the 4-th
order correlation function of the process $\hat\Omega({\bf
r}(t))$. Our calculations in Appendix \ref{AppB} show that it has
small numerical prefactor, $K_4(t)\simeq\-0.01 \cdot\eta^4(\nu
t)^3$, so that this cumulant becomes comparable to $K_2$ only at
rather long times. Thus, $K_2(t)$ dominates the polarization decay
at small times:
\begin{equation}\label{fca}
P(t)\approx e^{K_2(t)}= e^{-(t/t_S)^{3/2}}\!,\quad t_S=\frac 1\nu
\left(\frac{3\sqrt{2\pi}}{8\,\eta^2}\right)^{\frac23}.
\end{equation}
Excellent accuracy of this scaling at short times is seen from
comparison with the direct numerical simulations in
Fig.~\ref{scaling}. The similar decay law, $P(t)\sim
\exp{(-t^{3/2})}$, has been obtained in earlier studies, which
assumed the single-axis local hyperfine fields (directed along the
$z$-axis) \cite{CzK, Reineker, LeDoussal}, or the hyperfine fields
with fixed amplitude randomly distributed in the $x$--$y$ plane
\cite{RaikhSpinRel}. Our results confirm this decay law for the
hyperfine fields distributed isotropically in space, and show that
this feature holds for a very wide range of problems related to
the spin decay during 1D diffusion; we will also see the same
decay law below, for the vertical spin valve case.

More importantly, we notice that $K_2(t)$ and $K_4(t)$ depend only
on the single dimensionless renormalized time $\tau= (\nu t)
\eta^{4/3}$. From Eqs.~(\ref{MEm}) and (\ref{eqwrr}) one can see
that all cumulants, as well as $P(t)$ itself, are the functions of
two dimensionless quantities, $(\nu t)$ and $\eta$. However, our
analytical and numerical studies evidence a much stronger result,
that $P(t)$ is a function of a {\it single\/} dimensionless
quantity $\tau$. We performed a series of simulations for
different values of $\eta$, and Fig.~\ref{scaling} shows that all
results fall on the same universal curve $F(\tau)$, given in the
inset of Fig.~\ref{scaling}. This holds at all times we studied,
even at large $\tau$, when the contribution from the high-order
cumulants is important.

Even more, we see that the same scaling holds when we consider the
polarization decay at finite magnetic fields, finite electric
fields, as well as for the case of the vertical spin valve (both
without fields and with external magnetic and/or electric fields).
Thus, it is highly likely that the renormalized time $\tau$
represents a universal feature of the spin decay for $d=1$ random
walk. Understanding of this remarkable scaling, as far as we know,
is lacking.

Another interesting feature, seen from Fig. \ref{scaling}, is the
decay of $P(t)$ at long times, which has a stretched-exponential
form $P(t)\sim \exp\bigl(-\alpha\,\eta (\nu t)^{3/4}\bigr)$, with
$\alpha=1.33$; we checked that this form remains very accurate all
the way down to $P(t)\sim 10^{-12}$. Again, to our knowledge, the
reasons for this behavior are not understood yet.

Equally interesting is the behavior of the time-integrated
polarization $\sigma({\bf r})$. We are not aware of any analytical
theory, which would provide insights in the behavior of this
quantity and guide our simulations. Thus, we rely solely on the
numerical results.

Our numerical simulations show that in the whole range of
parameters $\sigma(r)$ has the exponential form, $\sigma(r)=l_S
\exp{(-|r|/l_S)}$. Without external fields, the time-integrated
polarization precisely follows the scaling law
$\sigma(r)=\eta^{-2/3} G\bigl(r\eta^{2/3} \bigr)$, where the
scaling function $G(w)=0.68\exp{(-1.47|w|)}$ is obtained from
numerical fitting. The origin of this scaling, as well as the
origin of the exponential dependence of $\sigma(r)$, are not
clear.

Note that the exponential decay of $\sigma(r)$ in the case of 1D
persistent diffusion is not trivial. If multiple returns were
negligible (as in transient diffusion in 3D), $P(t)$ would decay
exponentially with the decay time $t_{S,{\rm tr}}$; the space- and
time-integrated polarizations then would be related by the simple
convolution \cite{footnote}, $\sigma(\mathbf{r})=\nu \int_0^\infty
dt P(t) q_\mathbf{r}(t)$, where $q_\mathbf{r}(t)$ is the
probability to find the carrier at the site $\bf r$ at time $t$,
see Eq.~(\ref{MEp}). This would lead to the exponential decay
$\sigma(r)=\exp{(-|r|/l_{S,{\rm tr}})}$, with the well-known
diffusion relation $l_{S,{\rm tr}}=\sqrt{Dt_{S,{\rm tr}}}$, where
$D$ is the diffusion coefficient (for a random walk on a
$d$-dimensional lattice, $D=\nu a^2/2d$, where $a$ is the average
distance between the sites).

However, in our case, where the returns are crucial, $\sigma({\bf
r})$ and $P(t)$ are not related in such a simple way, and it is
not even clear whether such a relation exists. Indeed, if we used
the same convolution of the non-exponential $P(t)$ with
$q_\mathbf{r}(t)$ for 1D diffusion, we would obtain clearly
non-exponential decay law for $\sigma(r)$. Thus, the origin of the
exponential decay for $d=1$ must be different from that of $d=3$ case; this guess is
supported by the qualitative difference in response of $P(t)$ and
$\sigma(r)$ to the external magnetic and electric fields between the $d=1$ and $d=3$ cases.

\subsection{Role of external magnetic field}

\begin{figure}[t]
\vspace{-0.4cm}
\centerline{\includegraphics[width=95mm,angle=0,clip]{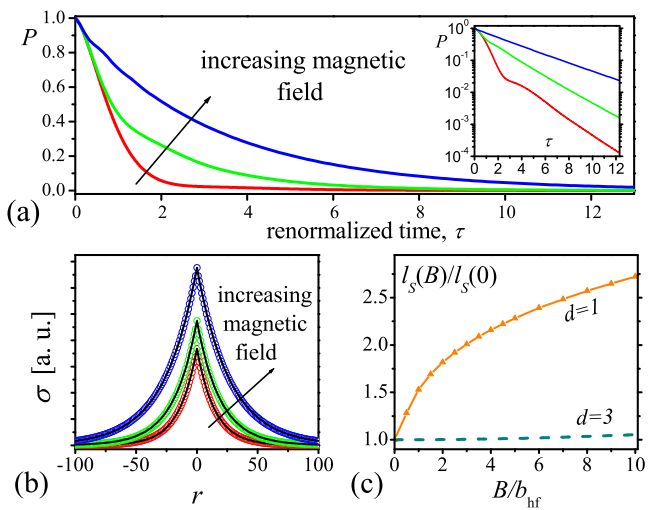}}
\caption{(Color online) Magnetic field dependence of spin
relaxation for a diffusion in $d=1$ with $\eta=0.01$. Red, green,
and blue plots correspond to the magnetic field values $B=0$,
$0.5\,b_{\text{hf}}$, and $2\,b_{\text{hf}}$, respectively. (a)
P(t) is plotted against $\tau=(\nu t)\eta^{4/3}$; the red curve is
the same as the one in the inset of Fig.~\ref{scaling}. Inset: the
same plot in the log-scale. (b) $\sigma(r)$ is plotted versus $r$.
Black curves are our exponential fits. (c) The normalized spin
diffusion length, $l_S(B)/l_S(0)$, is plotted vs
$B/b_{\text{hf}}$. Dashed line indicates $l_S(B)/l_S(0)$ without
returns, plotted from Eq.~(\ref{lstd}).} \label{bmplot}
\end{figure}

It has been noticed previously \cite{RaikhSpinRel} that in low
dimensional diffusion, in the presence of multiple returns, $P(t)$
is very sensitive to the external magnetic field. For the magnetic
field $\bf B$, the cumulant expansion of $P(t)$ can be carried out
after applying the rotating-frame transformation, ${\boldsymbol
\mu}(t)\to\exp(t\,{\hat\Omega}_B){\boldsymbol \mu} (t)$, where
${\hat\Omega}_B$ is the skew-symmetric matrix formed of $\bf B$
(see Appendix \ref{AppB}). Taking the external field as directed
along the $z$-axis, we find the second cumulant
\begin{equation}\label{mfsc}
K^B_2(t)= -2b_{\text{hf}}^2 \int\limits_0^t\!dt_1\!
\int\limits_0^{t_1}\!dt_2 \frac{\cos\bigl(B[t_1-t_2]\bigr)}
{\sqrt{2\pi\nu(t_1-t_2)}}.
\end{equation}
It is instructive to compare Eqs.~(\ref{mfsc}) and (\ref{seccum}):
the cosine term in the integrand is the only difference between
the cumulants $K_2$ for zero magnetic field and $K_2^B$ for finite
magnetic field. This term induces a cutoff for $t\gtrsim B^{-1}$,
reducing the integral significantly. This is somewhat similar to
motional narrowing: because of the external magnetic field, the
transversal components of the total field seen by ${\boldsymbol
\mu}$ average out on the timescale $B^{-1}$.

Comparing Eqs. (\ref{fca}) and (\ref{mfsc}), one can see that the
cutoff induced by the external magnetic field becomes important
for $B\sim t_S^{-1}\sim \eta^{1/3} b_{\text{hf}}$, which is even
smaller than $b_{\text{hf}}$. Our numerical results (Fig.~\ref{bmplot}) clearly verify
this sensitivity already at very low
fields. This behavior is in striking contrast with the transient
diffusion in 3D, where the spin relaxation time would scale as
$t_{S,{\rm tr}}\propto\bigl(1+(B/\nu)^2\bigr)$, meaning that the
magnetic field effects would be visible only at very large fields
$B\sim\nu \gg b_{\text{hf}}$.

The time-integrated spin polarization $\sigma(r)$ also exhibits
strong sensitivity to the external magnetic field. It still has
exponential form, $\sigma(r)\propto \exp{\bigl[-|r|/l_S(B)
\bigr]}$, but the spin decay length sensitively depends on $B$;
Fig.~\ref{bmplot}(b) illustrates this dependence for $\eta=0.01$.
The magnetic field dependence of the (normalized) spin decay
length, $l_S(B)/l_S(0)$, is plotted in Fig.~\ref{bmplot}(c).
Again, the analogy to the case of the transient 3D diffusion is
superficial; this point is demonstrated in more detail in
Appendix~\ref{AppA} [see Eq. (\ref{lstd})], where the transient
diffusion case is analyzed, and its qualitative difference with
our results for $d=1$ are emphasized.

\subsection{Role of external electric field}

If a drive voltage is applied to the spin valve (Fig.~\ref{latvert}), the resulting electric field $\mathbf{E} =
E\hat{\mathbf{x}}$ leads to a change in the hopping rates along
and against the field direction (forward and backward hopping
rates $\nu_f$ and $\nu_b$, see Fig.~\ref{chain}(b)). Utilizing the
Miller-Abrahams hopping model, \cite{M-A} we take $\nu_b=(\nu/2)
\exp(-\varepsilon)$, where $\varepsilon = eEa/k_BT$, and
$\nu_f=\nu/2$ (independent of $E$). Overall, this would lead to slower motion of the carrier,
implying slower changes of the random hyperfine field acting on it, and therefore
(as it happens in the motional narrowing scenario) would produce
faster decay of $P(t)$ with increasing $\varepsilon$. In the
regime of 3D transient diffusion (see Appendix \ref{AppA} for
details) this is the most important effect: $P(t)$ would decay
exponentially, and the decay time would decrease as
$(\nu_f+\nu_b)/\nu$.

\begin{figure}[t]
\vspace{-0.4cm}
\centerline{\includegraphics[width=95mm,angle=0,clip]{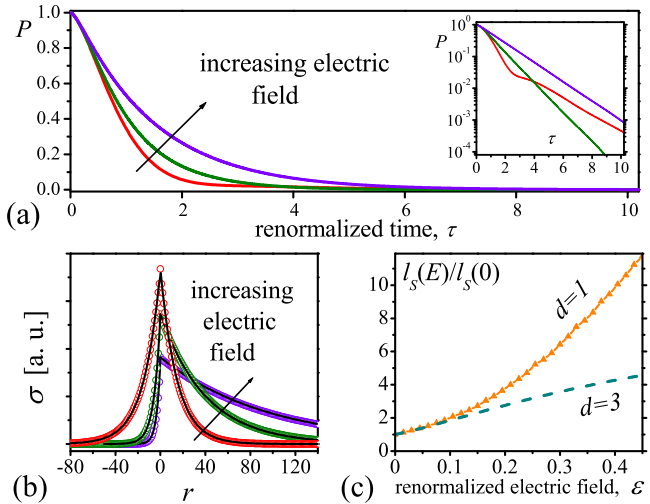}}
\caption{(Color online) Electric field dependence of spin
relaxation in $d=1$ with $\eta=0.01$, at zero magnetic field. With
typical $a\approx1$ nm and $T\approx100$ K, parameter $E\equiv
eEa/k_BT$ corresponds to the electric field $E\approx
8.6\,\varepsilon$~mV/nm. The red, dark green, and violet plots
correspond to $\varepsilon=0$, $0.15$, and $0.3$, respectively.
(a) P(t) is plotted versus $\tau=(\nu t)\eta^{4/3}$; the red curve
is the same as in Fig.~\ref{bmplot}(a). Inset: the same in the
log-scale. (b) $\sigma(r)$ is plotted versus $r$. Black lines are
our exponential fits. (c) Normalized spin diffusion length
$l_S(\varepsilon)/l_S(0)$ versus $\varepsilon$ (orange). The
dashed line indicates the same dependence without multiple
returns, plotted from Eq.~(\ref{lsvse}).} \label{efplot}
\end{figure}

However, in the persistent-diffusion regime, the returns are
important; besides inducing faster hopping, the electric field
also changes the statistics of the returns. By making the forward
hops more probable, the probability of the returns is decreased,
thus profoundly affecting the carrier spin relaxation. As above,
we calculate the short-time behavior of $P(t)$ using the cumulant
expansion, and the second cumulant in the presence of finite
electric field is
\begin{equation}\label{K2E}
K^E_2(t)= -b_{\text{hf}}^2\! \int\limits_0^t\!dt_1\!\!
\int\limits_0^{t_1}\!\!dt_2 \frac{e^{
-(\sqrt{\nu_f}-\sqrt{\nu_b})^2[t_1-t_2]}}
{(\nu_f\nu_b)^{1/4}\sqrt{\pi(t_1-t_2)}}.
\end{equation}
Here the exponent shows that the electric field
prevents multiple returns. Because of the exponent, the integral in
Eq.~(\ref{K2E}) mainly decreases with increasing $E$ (see Appendix \ref{AppB}
for more details). This means that the spin relaxation slows down with
increasing electric field, at least at relatively short times. The expected effect
is seen in Fig.~\ref{efplot}(a), which demonstrates simulation results for $P(t)$ for three
different values of $E$.

Now we turn to the time-integrated spin polarization, $\sigma(r)$.
Our numerical analysis shows that, like in all cases above, in the
presence of finite electric field
$\sigma(r)$ has exponential form, $\sigma(r)\propto
\exp{[-r/l_S(E)]}$; see an example in Fig.~\ref{efplot}(b) for
$\eta=0.01$. The curves are not symmetric with respect to $r=0$,
which is an obvious result of the drift induced by the electric
field. Also note that, in contrast with two previous cases (no
external fields and external magnetic field), the value of
$\sigma(0)$,
which sets the magnitude scale for the whole curve, is not
proportional to $l_S(E)$ anymore, and decreases with increasing
$E$, although $l_S(E)$ itself increases.

The spin decay length is very sensitive to the electric field: in
Fig.~\ref{efplot}(c) we show the dependence of $l_S(E)$ on the
normalized electric field $\varepsilon=eEa/k_BT$ for $\eta=0.01$,
where $l_S(0)\approx 14.9$. This is to be contrasted with the
similar dependence for a transiently diffusing carrier in $d=3$ with the
same $l_S(0)$, which is shown in the same graph for comparison:
without returns, $l_S(E)$ shows much weaker changes with electric
field.

\section{Spin relaxation in a vertical spin valve}

\label{SecVert}

The geometry of the vertical spin valve, Fig.~\ref{latvert}(b),
suggests diffusion over a linear chain of a finite length $L$.
Neglecting the back-tunneling into the electrodes, as it often
happens in experiments, we obtain the perfectly reflecting
boundaries. The spin relaxation now depends on the length of
the system: $P(t)= P(t,L)$, $\sigma(r)= \sigma(r,L)$. Another
feature of this geometry is that, instead of the whole function
$\sigma(r,L)$, one is interested in its value at the detection
electrode, $\sigma_\ast(L)\equiv\sigma(L,L)$.

\subsection{$P(t)$ and $\sigma_\ast(L)$ in the absence of external fields}

\begin{figure}[t]
\vspace{-0.4cm}
\centerline{\includegraphics[width=95mm,angle=0,clip]{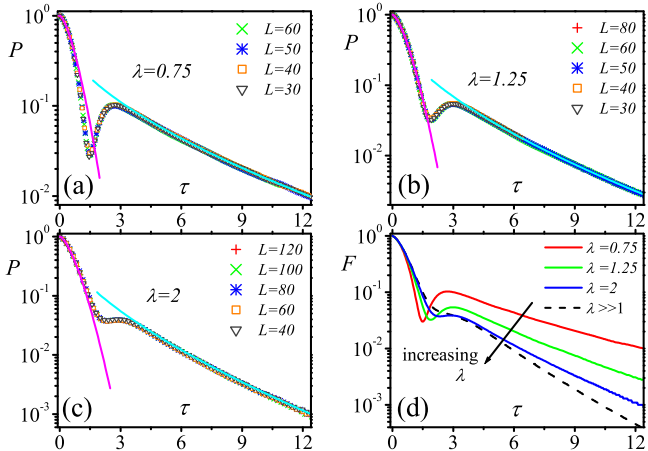}}
\vspace{-0.4cm} \caption{(Color online) Spin relaxation in a
finite chain of length $L$. When plotted versus $\tau=(\nu
t)\,\eta^{4/3}$ and $\lambda=L\eta^{2/3}$, including (a)
$\lambda=0.75$, (b) $\lambda=1.25$, and (c) $\lambda=2$, all
$P(t,L)$-points fall on universal curves. Cyan lines are stretched
exponential fits, $P(t,L)\sim\exp(-\alpha t^\beta)$, with (a)
$\beta=0.59$, (b) $\beta=0.61$, and (c) $\beta=0.64$. Magenta
lines are our theoretical fits with $\exp (-8\tau^{3/2}/3
\sqrt{\pi})$. (d) The resulting universal scaling function,
$F(\tau,\lambda)$, is plotted for the above values of $\lambda$.
The dotted line illustrates saturation of $F(\tau,\lambda)$ for
$\lambda\to\infty$.} \label{rflscl}
\end{figure}

Let a carrier be implanted at the boundary site, $r=1$, of a
linear chain of finite length $L$. After $n$ hops it will
diffusively cover the distance $\sim\sqrt{n}$. With the hopping
rate $\nu$ one has $n\approx\nu t$, so that for relatively short
times, $\nu t< L^2$, the boundary at $r=L$ will not affect the
spin relaxation noticeably. Therefore, for relatively short times,
$P(t)$ can be found by considering a carrier diffusing over the
semi-infinite chain, $r=1,2,...$, with the reflecting boundary at
$r=1$. Calculation carried out in Appendix \ref{AppB} for this
case gives the second cumulant function,
\begin{equation}\label{Kgcum}
K_2^{>}(t)\simeq-\frac {8\eta^2} {3\sqrt{\pi}}(\nu t)^{3/2},
\end{equation}
which differs from $K_2$ of the infinite chain,
Eq.~(\ref{seccum}), only by the factor $\sqrt{2}$. The ensuing
short-time superexponential dependence
of $P(t)$ is confirmed in our simulations, see Fig.~\ref{rflscl}.

At longer times, the influence of both boundaries becomes
noticeable, and the decay of the spin polarization will depend on
$L$. This dependence can be guessed using the results for
unbounded diffusion given above.

As we have seen in the previous Section, the characteristic length
of the spin relaxation is $l_S$, and scales with $\eta\equiv
b_{\text{hf}}/\nu$ as $\eta^{-2/3}$. Based on this fact and on
Eq.~(\ref{Kgcum}), we can guess that the spin polarization should
depend on the dimensionless time $\tau=(\nu t)\eta^{4/3}$ and on
the dimensionless length $\lambda=L\eta^{2/3}$. Our numerical
simulations confirm the expected  scaling law
$P(t,L)=F(\tau,\lambda)$, and provide the most notable features of
the scaling function $F(\tau,\lambda)$. In Fig.~\ref{rflscl}
(a)--(c), we demonstrate the scaling by plotting $P(t,L)$ as a
function of $\tau$ and $\lambda$, for three different values of
$\lambda$ and twelve different values of $\eta$.

These figures also show that the scaling function has the
superexponential form $F(\tau,\lambda) =\exp(-8\tau^{3/2}/3
\sqrt{\pi})$ at small times ($\tau\lesssim 0.5$), in accordance
with Eq.~(\ref{Kgcum}), and is independent of the normalized chain
length $\lambda$. At large times, the scaling function is
accurately described by the stretched exponential,
$F(\tau,\lambda)\sim \exp\bigl(-\alpha(\lambda)
\tau^{\beta(\lambda)}\bigr)$, with the parameters which depend on
the chain length; specifically, $\beta(0.75)=0.59$,
$\beta(1.25)=0.61$, and $\beta(2)=0.64$ in Figs. \ref{rflscl}
(a)-(c), respectively. As the length of the chain increases, the
exponent $\beta(\lambda)$ increases, saturating at the value
$\beta=0.75$ for very large $\lambda$ that corresponds to the
unbounded diffusion [see Fig.~\ref{rflscl}(d)].

In a similar way, we numerically verify the existence of scaling
for the time-integrated spin polarization. As we expect, all
length scales are scaled by the factor $\eta^{-2/3}$, so that
$\sigma(r,L)=\eta^{-2/3} G\bigl(r\eta^{2/3},L\eta^{2/3} \bigr)$,
in analogy with the case of unbounded diffusion in the lateral
spin valve. Correspondingly, for the spin polarization
$\sigma_\ast(L)$, observed at the detector electrode, we have the
scaling $\sigma_\ast(L)=\eta^{-2/3}G_\ast\bigl(L\eta^{2/3}
\bigr)$. The numerical fitting shows that the scaling function is
very accurately described as exponential,
$G_\ast(w)=2.712\exp(-1.475w)$. This corresponds to the
dependence, $\sigma_\ast(L)=4l_S \exp (-L/l_S)$, with the
diffusion length, $l_S=0.678\,\eta^{-2/3}$, which is nearly
identical to the one found for the unbounded diffusion in the
lateral spin valve.

\subsection{Role of external magnetic and electric fields}

In analogy with the case of the lateral spin valve, we studied the
influence of the magnetic field $B\hat{\bf z}$ along the $z$-axis,
and of the electric field $E\hat{\bf x}$ directed along the
$x$-axis, on the spin relaxation.

\begin{figure}[t]
\vspace{-0.3cm}
\centerline{\includegraphics[width=95mm,angle=0,clip]{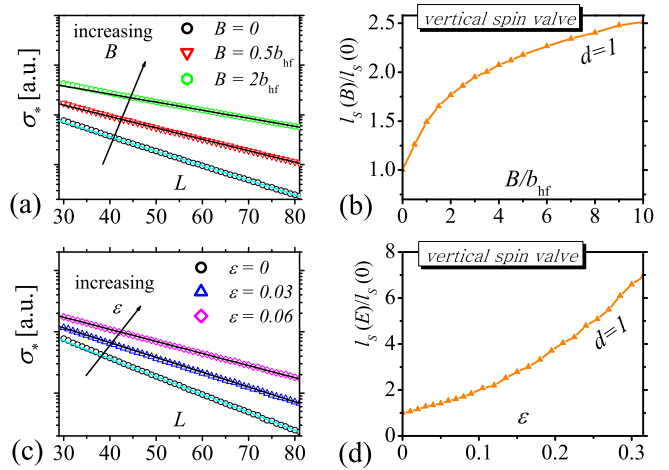}}
\vspace{-0.4cm} \caption{(Color online) Spin polarization at the
detection electrode, $\sigma_\ast(L)$, for a system with
$\eta=0.01$. (a) Log-plot of $\sigma_\ast(L)$ in zero electric
field and different magnetic fields: $B=0$ (black),
$0.5\,b_{\text{hf}}$ (red), $2\,b_{\text{hf}}$ (green). (c)
Log-plot of $\sigma_\ast(L)$ in three different electric fields,
$\varepsilon=0$ (black; the same as in (a)), $\varepsilon=0.03$
(blue), $\varepsilon=0.06$ (magenta), in zero magnetic field. The
black and cyan lines in (a) and (c) are our exponential fits. (b)
and (d): Dependence of the (normalized) diffusion length,
$l_S/l_S(0)$, on external magnetic and electric fields,
respectively. } \label{bmefrfl}
\end{figure}

We are primarily interested in the behavior of $\sigma_\ast(L)$.
Our simulations (Fig.~\ref{bmefrfl}) show that it remains
essentially exponential with $L$, in both external magnetic and
electric fields. Thus, the field dependence of $\sigma_\ast(L)$ is
fully encompassed by the field dependence of the spin decay length
on $B$ and $\varepsilon=eEa/k_BT$, shown in Figs.~\ref{bmefrfl}(b) and
(d). As before, we see that the returns lead to strong sensitivity
of $l_S$ to the external fields, and this dependence is very close
to its analog established in the previous Section. The curves in
Figs. \ref{bmefrfl}(b) and (d) appear to resemble the ones in Fig.
\ref{bmplot}(c) and Fig. \ref{efplot}(c).

\section{Spin Relaxation in $d=2$}

\label{SecTwoD}

\begin{figure}[t]
\vspace{-0.3cm}
\centerline{\includegraphics[width=95mm,angle=0,clip]{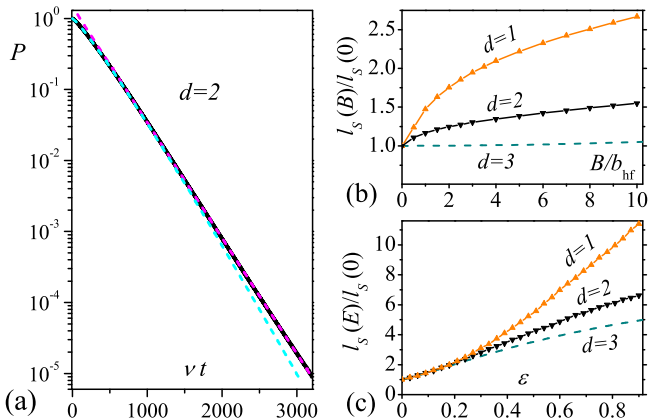}}
\vspace{-0.4cm} \caption{(Color online) Spin relaxation of  a
carrier diffusing over $d=2$ regular lattice, with $\eta\equiv
b_{\text{hf}}/\nu=0.025$. (a) Log-plot of $P(t)$ against $\nu t$
(black). The cyan dashed curve is plotted from Eq. (\ref{twodP}).
The magenta dashed line is our exponential fit. (b) Magnetic field
dependence of the (normalized) spin diffusion length; $l_S/l_S(0)$
is plotted vs $B$ for a diffusion in $d=1$ with $\eta_1=0.021$
(orange), $d=2$ with $\eta=0.025$ (black), and strong-collision
approximation to $d=3$ with $\eta_3=0.032$ (dashed line). (c)
Electric field dependence of the spin diffusion length.
$l_S/l_S(0)$ is plotted vs $\varepsilon = eEa/k_BT$, for a
diffusion in $d=1$ with $\eta_1=0.021$ (orange), $d=2$ with
$\eta=0.025$ (black), and $d=3$ with $\eta_3=0.032$ in
strong-collision approximation (dashed line).} \label{twodplot}
\end{figure}

Above we focused on the analysis of the spin relaxation of a
diffusing carrier for $d=1$, mainly because the effect of multiple
returns is strongest in this case. Meanwhile, it is rather
straightforward to extend our analysis to the $d=2$ case. For a
carrier performing a simple random walk over a regular lattice in
$d=2$, the second cumulant function is calculated in Appendix
\ref{AppB}. It suggests the short-time decay
\begin{equation}\label{twodP}
P(t)\simeq\exp \bigl[-2\,\eta^2(\nu t)\ln(\gamma\nu t)/\pi\bigr],
\end{equation}
where $\gamma=5.243$ is a numerical coefficient. We have checked
numerically that this formula accurately describes the spin
relaxation down to rather small polarization values. Specifically,
at $P(t)\sim0.05$, the observed deviation from Eq. (\ref{twodP})
was only about $0.002$. Our simulations have also shown that at
longer times the decay slows down, closely resembling exponential.
This is shown in Fig.~\ref{twodplot}(a), which illustrates the
spin relaxation in a system with $\eta=0.025$. Thus, in $d=2$ the
effect of multiple returns is quite noticeable. Meanwhile, it is
rather easy to check that in $d=3$ the polarization decay is
exponential at virtually all times.

To understand the spin-transport relaxation in $d=2$, we consider
a lateral spin valve device similar to that of
Fig.~\ref{latvert}(a), where the carriers hop between the sites of
$d=2$ regular lattice located in the $x$--$z$ plane [see
Fig.~\ref{chain}(c)], so that each site is characterized by the
radius vector $\mathbf{r} =(x,z)$, with $x,z=0, \pm 1,\pm 2,..$.
Assuming that the size of the organic layer in
$\hat{\mathbf{z}}$-direction is much larger than the spin
diffusion length, we characterize the spin-transport relaxation by
the quantity,
\begin{equation}\label{sig2d}
\sigma_{2d}(x)= \sum_{{\bf r}=(x,z)\atop x=L} \sigma(\mathbf{r})
\equiv \nu \sum_{{\bf r}=(x,z)\atop x=L} \int_0^\infty dt\langle
m_\mathbf{r}^z (t)\rangle_{\text{hf}},
\end{equation}
which is the total time-integrated spin polarization that has
reached the detection electrode at $x=L$. Here
$\mathbf{m}_\mathbf{r} (t)$ is the solution of drift-diffusion
equation (\ref{MEm}) for the 2D regular lattice; we solve this
equation numerically, with the initial condition
$\mathbf{m}_\mathbf{r} (0)= \delta_ {\mathbf{r}, 0}
\mathbf{m}(0)$. The influence of the external magnetic and
electric fields is taken into account as described above.

Our results show that the decay of $\sigma_{2d}(x)$ is
exponential, both with and without the external fields. The spin
diffusion length $l_S$ is also rather sensitive to the external
fields, as a result of the returns in the course of the 2D random
walk. This is illustrated in Figs.~\ref{twodplot}(b) and (c),
where we compare the results for the random walk in different
dimensions for $\eta=0.025$, where the spin diffusion length in
the absence of the external fields is $l_S=9.0$. From the results
of Section \ref{SecLat} one can see that the same zero-field spin
diffusion length in $d=1$ corresponds to $\eta_1\approx 0.021$,
whereas for $d=3$, from the strong collision approximation (i.e.,
by neglecting multiple returns) we find the same zero-field spin
diffusion length for $\eta_3\approx 0.032$ [cf. Eq.~(\ref{lstd})].
Therefore, in Figs.~\ref{twodplot}(b) and (c) we compare the field
dependence of the normalized spin diffusion length, $l_S(B)/l_S$
and $l_S(E)/l_S$, for $d=1$ with $\eta_1=0.021$, $d=2$ with
$\eta=0.025$, and $d=3$ with $\eta_3=0.032$. Evidently, more
frequently occurring multiple returns lead to stronger growth of
spin diffusion length with external fields.

\section{Conclusion}

We have investigated spin relaxation of a carrier performing a
random walk on a lattice, with random magnetic fields at each
site. This models spin relaxation in organic semiconductors, where
the charge transport between the $\pi$-conjugated segments of
molecules is incoherent, and where the carrier spin interacts with the hydrogen nuclear spins surrounding a
segment. Due to relatively large number of surrounding
nuclear spins and their slow dynamics, the on-site random magnetic
fields are taken as static, and sampled from the Gaussian distribution. The
width of the distribution reflects the strength of hyperfine
coupling at each site. For the incoherent hopping we
assumed a random walk with the constant transition rates.

To understand the effect of multiple self-intersections of random
walks, we have focused on the motion in $d=1$ dimensional infinite
linear chain. A superexponential short-time decay of spin
polarization, $P(t)\sim\exp(-\alpha_1 t^{3/2})$, was found
analytically from the cumulant expansion. The
numerical simulations confirmed the superexponential dependence, and showed that it changes to a stretched exponential at longer times. We have also analyzed the
spin relaxation of a carrier diffusing over a linear chain of
finite length, and established that the short-time
relaxation is somewhat faster, $P(t)\sim\exp(-\sqrt{2}\alpha_1
t^{3/2})$, whereas the stretched exponential long-time decay
becomes slower. As a consequence of the multiple returns, in all these
cases $P(t)$ is highly sensitive to the external magnetic and electric
fields. For a diffusion over $d=2$ regular lattice, we have found
the short-time behavior $P(t)\sim\exp(-\alpha_2 t\ln t)$, smoothly
crossing over to the exponential long-time decay.

Due to its relevance for the spin transport experiments, we have
investigated the spin relaxation in the space domain, considering
the time-integrated spin polarization $\sigma({\bf r})$. It was
demonstrated that, despite the strongly non-exponential decay in
time, the spin relaxation in space is essentially exponential,
superficially similar to that of a carrier diffusing in $d=3$.
However, diffusion in lower dimensions shows much stronger sensitivity to the external electric and magnetic fields.

\section*{Acknowledgments}

We thank J. Shinar and M. E. Raikh for many useful discussions.
Work at the Ames Laboratory was supported by the US Department
of Energy, Office of Science, Basic Energy Sciences, Division
of Materials Sciences and Engineering. The Ames Laboratory
is operated for the US Department of Energy by Iowa State
University under Contract No. DE-AC02-07CH11358.


\appendix

\section{}

\label{AppA}

In this Appendix we calculate spin relaxation for a carrier
diffusing transiently, i.e., when self-intersections of its
random-walk trajectories are negligible, and the strong-collision
approximation \cite{Kubo79} is valid. Our starting point is Eq.
(\ref{dbav}) of the main text. Let $\mathbf{r}(t)$ be a random
walk trajectory that starts at $\mathbf{r}_0=0$ and passes through
$n$ sites, $\mathbf{r}_1,..,\mathbf{r}_n$. Then the time-ordered
exponent can be written as
\begin{equation}\label{prodexp}
T\exp\int_0^tdt'{\boldsymbol \Omega}(\mathbf{r}(t'))=
e^{\tau_n{\boldsymbol \Omega}(\mathbf{r}_n)} \cdots
e^{\tau_0{\boldsymbol \Omega}(\mathbf{r}_0)},
\end{equation}
where $\tau_k$ are the waiting time at $\mathbf{r}_k$,
$k=0,1,..,n$. In the absence of self-intersections all
$\mathbf{r}_k$ are different, and the average of Eq.
(\ref{prodexp}) over the local hyperfine frequencies is a product
of exponents, averaged over the independent Gaussian distributions
of hyperfine frequencies ${\{{\bf b}_{\mathbf{r}_k}\}}$, namely
\begin{equation}\label{avrgd}
\Bigl\langle T\exp\int_0^tdt'{\boldsymbol \Omega} (\mathbf{r}
(t'))\Bigr\rangle _{\text{hf}}{\Big |}_{zz} =
\prod_{k=0}^nf(\tau_k),
\end{equation}
where $f(\tau)$ is defined by $\langle e^{\tau{\boldsymbol
\Omega}_{\bf b}} \rangle _{\{{\bf b}\}} =\hat{\mathbf{1}}f(\tau)$,
with ${\boldsymbol \Omega}_{\bf b}$ being the skew-symmetric
matrix formed of $\bf b$. By averaging
\begin{equation}\label{expexp}
e^{\tau{\boldsymbol \Omega}_{\bf b}} = \hat{\mathbf{1}}+
\sin(|{\bf b}|\tau)\frac{{\boldsymbol \Omega}_{\bf b}}{|{\bf b}|}
+2\sin^2(|{\bf b}|\tau/2) \frac{{\boldsymbol \Omega}_{\bf
b}^2}{|{\bf b}|^2},
\end{equation}
over the Gaussian distribution of $\boldsymbol  b$ with zero mean
and standard deviation, $b_{\text{hf}}$, one gets:
\begin{equation}\label{KT0}
f(\tau)=\frac13 + \frac23\bigl(1-b_{\text{hf}}^2\tau^2\bigr)
\exp\bigl(-b_{\text{hf}}^2\tau^2/2\bigr).
\end{equation}
When a longitudinal magnetic field ${\bf B}=B\hat{\mathbf{z}}$ is
applied,
Eq. (\ref{expexp}) should be averaged over a Gaussian distribution
of $\boldsymbol  b$ with the mean, $\langle{\bf b}\rangle=
 B\hat{\mathbf{z}}$. Even though this leads to a
non-diagonal matrix $\langle e^{\tau{\boldsymbol \Omega}_{\bf b}}
\rangle _{\{{\bf b}\}}$, it remains block-diagonal, so that Eq.
(\ref{avrgd}) holds with a modified $f(\tau)$. Note in passing
that $f(\tau)$ is a typical example of the static Kubo-Toyabe
relaxation function. \cite{Kubo79, KuboToyabe}

Next we want to average Eq. (\ref{avrgd}) over the waiting time
distributions and random-walk trajectories. Because of the absence
of returns, the latter reduces to a summation over all $n$, whereas
the former can be done by integrating $f(\tau)$
with the waiting-time distribution function, $\nu\, e^{-\nu\tau}$.
Hence from Eqs. (\ref{dbav}) and (\ref{avrgd}) we get:
\begin{eqnarray}\label{Pttr}
P(t)=\sum_{n=0}^\infty\int\limits_0^\infty\!\! d\tau_0\cdots\!
\int\limits_0^\infty\!\! d\tau_n\prod_{j=0}^n
f(\tau_j) \nu e^{-\nu\tau_j}&&
\nonumber\\
\times \left[\theta\biggr(t-\sum_{k=0}^{n-1}\tau_k\biggl)
-\theta\biggr(t-\sum_{k=0}^n\tau_k\biggl) \right]\!.&&
\end{eqnarray}
Here, the difference of $\theta$-functions guarantees that at time
$t$ the walker has performed exactly $n$ steps, so that
$\sum_{k=0}^{n-1}\tau_k < t < \sum_{k=0}^n\tau_k$. Using the
integral representation,
$\theta(x)=\int\bigl[e^{izx}/(z-i\epsilon)\bigr] dz/(2\pi i)$, we
reduce Eq. (\ref{Pttr}) to
\begin{equation}\label{Ptsimp}
P(t)=\int\limits_{-\infty}^\infty\!\!\frac{dz}{2\pi i}\frac{e^{iz\nu t}}
{z-i\epsilon}\frac{u(1)-u(1+iz)}{1-u(1+iz)},
\end{equation}
where $u(y)=\nu\int_0^\infty\!\!d\tau
f(\tau)e^{-y\nu\tau}$. Going back to the
definition of $f(\tau)$ ant taking the $\tau$-integral we get:
\begin{equation}\label{uyir}
u(y)=\frac1y\int\!\! d^3{\boldsymbol \zeta}\,
\frac{e^{-\frac{|{\boldsymbol \zeta}|^2}2}} {(2\pi)^{\frac32}}
\frac{(y\eta)^2+(\zeta_z+\beta)^2}
{(y\eta)^2+\zeta_r^2+\zeta_y^2+(\zeta_z+\beta)^2},
\end{equation}
where $\beta = B/b_{\text{hf}}$. Exact evaluation of this integral
yields $u(y)$ in terms of the error function. However, we are
interested in small values of $\eta\equiv b_{\text{hf}}/\nu$,
where the most relevant pole of Eq. (\ref{Ptsimp}), given by $u(1+
iz_0)=1$, is located at $|z_0|\ll1$. Therefore $z_0$ can be found
from the small-$\eta$ ($\eta\ll\beta^{-1}$, $1$)
expansion of Eq. (\ref{uyir}) for $|y|\sim 1$,
\begin{equation}\label{uyas}
u(y)\simeq\frac1y\left(1- \frac2 {(y/\eta)^2+ \beta^2 +1}\right),
\end{equation}
which in fact provides a good approximation for any $\beta$. Form
Eq. (\ref{uyas}) we find $z_0\approx 2i/(\eta^{-2}+\beta^2+1)$,
yielding
\begin{equation}\label{Ptans}
P(t)\simeq e^{-t/t'_S},\quad t'_S=(\eta^{-2}+\beta^2+1)/2\nu.
\end{equation}

The spacial dependence of spin polarization is given by
\begin{eqnarray}
\label{sgr} \sigma(\mathbf{r}) =&&\!\!\!\!\!
\nu\!\int\limits_0^\infty\!\! dt \sum_{n=0}^\infty \!
Q_n(\mathbf{r})\!\! \int\limits_0^\infty\!\! d\tau_0\cdots\!\!
\int\limits_0^\infty\!\! d\tau_n\!
\prod_{j=0}^n\!f(\tau_j) \nu e^{-\nu\tau_j}
\nonumber\\
&&\times \left[\theta\biggr(t-\sum_{k=0}^{n-1}\tau_k\biggl)
-\theta\biggr(t-\sum_{k=0}^n\tau_k\biggl) \right]\!,
\end{eqnarray}
where $Q_n(\mathbf{r})$ is the probability that the random walker
is at $\mathbf{r}$ after $n$ steps. Calculating the integrals in
Eq. (\ref{sgr}) is easy by taking first the $t$-integral.
Further, introducing $\tilde{u}=\nu^2\int_0^\infty\!\!d\tau \tau
f(\tau)e^{-\nu\tau}$ and using Eq.
(\ref{uyas}), we find:
\begin{equation}\label{sigres}
\sigma(\mathbf{r}) =\tilde{u}\sum_{n=0}^\infty Q_n(\mathbf{r})
\bigl[u(1)\bigr]^n \simeq\sum_{n=0}^\infty
Q_n(\mathbf{r})e^{-n/\nu t'_S}.
\end{equation}
The probability $Q_n(\mathbf{r})$ is related to the solution of
Eq.~(\ref{MEp}) as $q_{\mathbf{r}}(t)=e^{-\nu t}\sum_{n=0}^\infty
Q_n(\mathbf{r})(\nu t)^n/n!$. This can be used in Eq. (\ref{sigres})
to express $\sigma(\mathbf{r})$ in terms of $q_{\mathbf{r}}(t)$:
\begin{equation}\label{svq}
\sigma(\mathbf{r})\simeq\nu\int_0^\infty \!\!\!dt\,
q_{\mathbf{r}}(t) P(t).
\end{equation}
The large-$r$ behavior of $\sigma(\mathbf{r})$ follows from that
of $q_{\mathbf{r}}(t)$. Namely, for a simple random walk on a
$d$-dimensional regular lattice one has
$q_{\mathbf{r}}(t)=(2\pi\nu t/d)^{-d/2}
\exp\bigl[-d|\mathbf{r}|^2/(2\nu t)\bigr]$, leading to the
exponential decay,
$\sigma(\mathbf{r})\propto\exp\bigl(-|\mathbf{r}|/l_S\bigr)$, with
\begin{equation}\label{lstd}
l_S(B)=\sqrt{\frac{\nu t'_S(B)}{2d}} \simeq
\frac1{\sqrt{4d}\,\eta} \sqrt{1 + \bigl( \eta B/
b_{\text{hf}}\bigr)^2}.
\end{equation}

In an external electric field, $\mathbf{E}=E\hat{\mathbf{x}}$,
hopping rates along $\hat{\mathbf{x}}$ are changed. This leads to
the drift along $\hat{\mathbf{x}}$, and also modifies the
diffusion and the waiting-time distribution. While $P(t)$ is
affected only because of the change in waiting-time distribution,
$q_{\mathbf{r}}(t)$ and consequently $\sigma(\mathbf{r})$ are
sensitive to the drift and diffusion. Assuming a random walk over
a $d$-dimensional regular lattice, and for the hopping model
considered in the main text, the hopping rates forward and
backward to $\hat{\mathbf{x}}$ are $\nu_f=\nu/2d$ and $\nu_b=\nu
e^{-\varepsilon}/2d$, where $\varepsilon= eEa/k_BT$, whereas in
perpendicular directions hopping rates are $\nu/2d$. From the
corresponding waiting-time distribution function,
$\tilde{\nu}e^{-\tilde{\nu}\tau}$ with $\tilde{\nu}=\nu_f+\nu_b
+\nu(d-1)/d$, one finds $P(t)\simeq \exp\bigl[-t/t'_S(\varepsilon)
\bigr]$ with the electric-field dependent spin relaxation time,
\begin{equation}\label{tausvse}
t'_S(\varepsilon)=\frac1{2\nu\,\eta^2} \frac {2d-1 +
e^{-\varepsilon}}{2d} +\frac{\beta^2+1}\nu \frac
d{2d-1+e^{-\varepsilon}}.
\end{equation}
The drift-diffusion equation (\ref{MEp}) in $d=3$ dimensions has
the solution,
\begin{equation}\label{qrd}
q_{\mathbf{r}}(t)=\left(\frac{\nu_f}{\nu_b}\right)^{\frac x2}
e^{-\tilde{\nu}t} I_x\bigl(2\sqrt{\nu_f\nu_b}\, t\bigr)
I_y\bigl(\nu t/d\bigr)I_z\bigl(\nu t/d\bigr),
\end{equation}
where $\mathbf{r} = (x,y,z)$ with $x,y,z=0,\pm1,\pm2,..$, and
$I_\alpha$ is the modified Bessel function of order $\alpha$. For
a spin valve similar to those illustrated in Fig. \ref{latvert},
we evaluate the quantity, $\sigma(x)=\sum_{y,z}
\sigma(\mathbf{r})$ (in lower dimensions, one or both of last
terms in Eq. (\ref{qrd}) should be eliminated, and the sum for
$\sigma(x)$ should be changed correspondingly). After taking this
sum, the integral Eq. (\ref{svq}) reduced to the Laplace transform
for a modified Bessel function, yielding $\sigma(x)\propto
\exp\bigl[-x/l_S(\varepsilon)\bigr]$ for all $x>0$, where
\begin{widetext}
\begin{equation}\label{lsvse}
l_S(\varepsilon)=\frac1{\ln\left(\frac12(1+e^{-\varepsilon}) +
\frac d{\nu t'_S(\varepsilon)} + \sqrt{\left[
\frac12(1+e^{-\varepsilon}) + \frac d{\nu t'_S(\varepsilon)}
\right]^2-e^{-\varepsilon}}\right)}.
\end{equation}
This dependence is plotted in Fig.~\ref{efplot}(c) for a system
with $\beta=0$ and $l_S(0)=14.9$ (corresponding to $\eta=0.0336$
in $d=1$), and in Fig.~\ref{twodplot}(c) for a system with
$\beta=0$ and $l_S(0)=9$ (corresponding to $\eta=0.0322$ in
$d=3$).

\section{}

\label{AppB}

In this Appendix we calculate the  $d=1$ dimensional second
cumulant functions, $K_2$, $K^B_2$, $K_2^E$, $K_2^{>}$, Eqs.
(\ref{seccum}), (\ref{mfsc}), (\ref{K2E}), and (\ref{Kgcum}), the
fourth cumulant function $K_4$, as well as
the second cumulant for a simple random walk on a two-dimensional
regular lattice, $K_2^{(2)}$. Basic ingredients of this
calculation are the Markov property of random walk and its Greens
function, $G(\mathbf{r}, \mathbf{r}',t)$, which is the solution of
corresponding random walk equation (\ref{MEp}) with the initial
condition, $q_{\mathbf{r}'}(0)=\delta_{\mathbf{r}, \mathbf{r}'}$.
The second cumulant function of Eq. (\ref{dbav})
is defined by
\begin{equation}\label{K2def}
K_2(t)=\int\limits_0^t\!dt_1\! \int\limits_0^{t_1}\!dt_2 \langle
\langle {\boldsymbol \Omega} (t_1)\, {\boldsymbol \Omega}
(t_2)\rangle \rangle _{zz},\qquad \langle \langle {\boldsymbol
\Omega} (t_1)\, {\boldsymbol \Omega} (t_2)\rangle \rangle \equiv
\bigl\langle \bigl\langle{\boldsymbol \Omega} (\mathbf{r}(t_1))\,
{\boldsymbol \Omega} (\mathbf{r}(t_2))\bigr\rangle
_{\text{hf}}\bigr\rangle _{\text{rw}},
\end{equation}
where the average over random walk trajectories and locally
Gaussian hyperfine frequencies is meant. Using the matrix form,
Eq. (\ref{ssm}), and the fact that the components of ${\bf
b}_\mathbf{r}$ are delta-correlated, $\langle
b_\mathbf{r}^\alpha\,  b_{\mathbf{r}'}^\beta \rangle _{\text{hf}}
=-b_{\text{hf}}^2\delta_{\mathbf{r},\mathbf{r}'} \delta
_{\alpha,\beta}$, one easily takes the average over local
frequencies, resulting in $\langle \langle {\boldsymbol \Omega}
(t_1)\, {\boldsymbol \Omega} (t_2)\rangle \rangle =-
\hat{\mathbf{1}}
\cdot2\,b_{\text{hf}}^2\,\langle\delta_{\mathbf{r}(t_1),
\mathbf{r}(t_2)}\bigr\rangle_{\text{rw}}$. For any type of random
walk, this can be expressed via the Greens function as follows:
\begin{equation}\label{viaG}
\langle \langle {\boldsymbol \Omega} (t_1)\, {\boldsymbol \Omega}
(t_2)\rangle \rangle =- \hat{\mathbf{1}}\cdot2\,b_{\text{hf}}^2\,
\sum_\mathbf{r} G(\mathbf{r},\mathbf{r},t_1-t_2)
G(0,\mathbf{r},t_2).
\end{equation}

For the random walk on an infinite chain we have
$G(r,r',t)=e^{-\nu t}I_{|r-r'|}(\nu t)$, where $I_r(z)$ is the
modified Bessel function of order $r$. From Eq. (\ref{viaG}) we
find, $\langle \langle {\boldsymbol \Omega} (t_1)\, {\boldsymbol
\Omega} (t_2)\rangle \rangle _{zz}=-2\, b_{\text{hf}}^2
e^{-\nu(t_1-t_2)}I_0\bigl(\nu(t_1-t_2)\bigr)$, which gives the
second cumulant,
\begin{equation}\label{K2ans}
K_2(t)=-2\,\eta^2 \int\limits_0^{\nu t}\!dz_1\!
\int\limits_0^{z_1}\!dz_2e^{-(z_1-z_2)}I_0(z_1 - z_2).
\end{equation}
In view of large $\eta$, it is necessary to find the integral for
large $(\nu t)\gg1$. Utilizing the large-$z$ asymptote
$e^{-z}I_r(z)\simeq (2\pi z)^{-1/2} \exp\bigl(-r^2/2z\bigr)$ in
the integrand, we arrive at the result Eq. (\ref{seccum}).

As the odd cumulants are zero, the fourth cumulant function
is expressed in terms of the four-time correlation function as follows:
\begin{equation}\label{K4def}
\frac 12 K_2^2(t)+ K_4(t)=\int\limits_0^t\!\!dt_1\!\!
\int\limits_0^{t_1}\!\!dt_2\!\! \int\limits_0^{t_2}\!\!dt_3\!\!
\int\limits_0^{t_3}\!\!dt_4 \langle \langle {\boldsymbol \Omega}
(t_1)\, {\boldsymbol \Omega} (t_2)\, {\boldsymbol \Omega} (t_3)\,
{\boldsymbol \Omega} (t_4) \rangle \rangle_{zz}.
\end{equation}
From Eq. (\ref{ssm}) we find that the $zz$-component of the
product ${\boldsymbol \Omega} (t_1)\, {\boldsymbol \Omega} (t_2)\,
{\boldsymbol \Omega} (t_3)\, {\boldsymbol \Omega} (t_4)$ is equal
to
\begin{equation}\label{4matr}
 b_{\mathbf{r}(t_2)}^z  b_{\mathbf{r}(t_3)}^z \Bigl[
 b_{\mathbf{r}(t_1)}^x  b_{\mathbf{r}(t_4)}^x +
 b_{\mathbf{r}(t_1)}^y  b_{\mathbf{r}(t_4)}^y\Bigr]
+\Bigl[ b_{\mathbf{r}(t_1)}^x  b_{\mathbf{r}(t_2)}^x +
b_{\mathbf{r}(t_1)}^y  b_{\mathbf{r}(t_2)}^y\Bigr]\Bigl[
 b_{\mathbf{r}(t_3)}^x  b_{\mathbf{r}(t_4)}^x +
 b_{\mathbf{r}(t_3)}^y  b_{\mathbf{r}(t_4)}^y\Bigr].
\end{equation}
To find $ \langle {\boldsymbol \Omega} (t_1)\, {\boldsymbol
\Omega} (t_2)\, {\boldsymbol \Omega} (t_3)\, {\boldsymbol
\Omega}(t_4)\rangle_{zz}$, we first average Eq. (\ref{4matr}) over
the local hyperfine field distribution, then over the random walk
trajectories. After the first averaging we get:
\begin{equation}\label{rwav} \langle {\boldsymbol \Omega}
(t_1)\, {\boldsymbol \Omega} (t_2)\, {\boldsymbol \Omega} (t_3)\,
{\boldsymbol \Omega} (t_4)\rangle_{zz}= b_{\text{hf}}^4\, \langle
4\,\delta_{\mathbf{r}(t_1),\mathbf{r}(t_2)}
\delta_{\mathbf{r}(t_3),\mathbf{r}(t_4)} +
2\,\delta_{\mathbf{r}(t_1),\mathbf{r}(t_3)}
\delta_{\mathbf{r}(t_2),\mathbf{r}(t_4)} +
4\,\delta_{\mathbf{r}(t_1),\mathbf{r}(t_4)}
\delta_{\mathbf{r}(t_2),\mathbf{r}(t_3)} \bigr\rangle
_{\text{rw}}.
\end{equation}
This equation follows from the calculation of local field averages
of the form, $\bigl \langle  b_{\mathbf{r}_1}^\alpha
 b_{\mathbf{r}_2}^\alpha  b_{\mathbf{r}_3}^\beta
 b_{\mathbf{r}_4}^\beta \bigr\rangle _{\text{hf}}$. For
$\alpha \neq \beta$, this calculation is simple and gives $
b_{\text{hf}}^4 \,\delta_{\mathbf{r}_1, \mathbf{r}_2}
\delta_{\mathbf{r}_3, \mathbf{r}_4}$. For $\alpha =\beta$, on the
other hand, it results in the combination, $
b_{\text{hf}}^4\bigl(\delta_{\mathbf{r}_1, \mathbf{r}_2}
\delta_{\mathbf{r}_3, \mathbf{r}_4} + \delta_{\mathbf{r}_1,
\mathbf{r}_3} \delta_{\mathbf{r}_2, \mathbf{r}_4}
+\delta_{\mathbf{r}_1, \mathbf{r}_4} \delta_{\mathbf{r}_2,
\mathbf{r}_3}\bigr)$. Note that contributions with $\mathbf{r}_1 =
\mathbf{r}_2 = \mathbf{r}_3 =\mathbf{r}_4$ cancel out from this
combination due to the Gaussian character of local frequency
distributions.

Next we average Eq. (\ref{rwav}) over the random walk
trajectories. For a function of four coordinates, $f$, and times
arranged as $t_1\geq t_2\geq t_3\geq t_4$, from the Markov property
of random walk one generally has:
\begin{equation}\label{f4}
\langle f\bigl(\mathbf{r}(t_1), \mathbf{r}(t_2), \mathbf{r}(t_3),
\mathbf{r}(t_4) \bigr) \bigr\rangle _{\text{rw}}=
\sum_{\mathbf{r}_1, .., \mathbf{r}_4}f(\mathbf{r}_1, \mathbf{r}_2,
\mathbf{r}_3, \mathbf{r}_4) G(\mathbf{r}_{12},t_{12})
G(\mathbf{r}_{23},t_{23}) G(\mathbf{r}_{34},t_{34})
G(\mathbf{r}_4,t_4),
\end{equation}
where $\mathbf{r}_{ij}= \mathbf{r}_i-\mathbf{r}_j$ and
$t_{ij}=t_i-t_j$. We apply Eq. (\ref{f4}) with the infinite-chain
Greens function to each term of Eq.~(\ref{rwav}), and find the
large- $\nu t_{ij}$ asymptotes of the resulting quantities:
\begin{eqnarray}
&&\langle\delta_{r(t_1),r(t_2)} \delta_{r(t_3),r(t_4)}\bigr\rangle
_{\text{rw}}= G(0,t_{12})G(0,t_{34})\simeq \frac
1{2\pi\nu\sqrt{t_{12}t_{34}}}, \label{rwav1}\\
&&\langle\delta_{r(t_1),r(t_3)} \delta_{r(t_2),r(t_4)}\bigr\rangle
_{\text{rw}}= \sum _r G(r,t_{12})G(r,t_{23})G(r,t_{34})\simeq
\frac 1{2\pi\nu\sqrt{t_{13}t_{24}-t_{23}^2}},\label{rwav2}\\
&&\langle\delta_{r(t_1),r(t_4)} \delta_{r(t_2),r(t_3)}\bigr\rangle
_{\text{rw}}= G(0,t_{23})\sum _r G(r,t_{12})G(r,t_{34})\simeq
\frac 1{2\pi\nu\sqrt{t_{14}t_{23}-t_{23}^2}}. \label{rwav3}
\end{eqnarray}
Equations (\ref{rwav}) and (\ref{rwav1})-(\ref{rwav3}) define the
integrand of Eq. (\ref{K4def}). Using the asymptotic forms, we
find:
\begin{equation}\label{ffint}
\int\limits_0^t\!\!dt_1\!\! \int\limits_0^{t_1}\!\!dt_2\!\!
\int\limits_0^{t_2}\!\!dt_3\!\! \int\limits_0^{t_3}\!\!dt_4
\langle \langle {\boldsymbol \Omega} (t_1)\, {\boldsymbol \Omega}
(t_2)\, {\boldsymbol \Omega} (t_3)\, {\boldsymbol \Omega}
(t_4)\rangle \rangle_{zz} \simeq\frac 59\eta^4(\nu t)^3.
\end{equation}
\end{widetext}
This relation, together with Eq. (\ref{K4def}), leads to the
result,
\begin{equation}\label{frthcum}
K_4(t)\simeq\bigl[( 5\pi-16)/9\pi\bigr]\eta^4(\nu t)^3.
\end{equation}
It is worthwhile to notice that Eqs. (\ref{rwav2}) and
(\ref{rwav3}) make the non-Gaussian character of the stochastic
process $\{{\boldsymbol \Omega}(t)\}$ explicit. Indeed, a Gaussian
$\{{\boldsymbol \Omega}(t)\}$ would entail a factorization of
four-time functions, as it happens in Eq. (\ref{rwav1}), whereas
Eqs. (\ref{rwav2}) and (\ref{rwav3}) do not satisfy this
condition.

Consider now the spin polarization decay in $d=1$, in the presence
of a magnetic field along $\hat{\mathbf{z}}$. This case can be
described by adding in Eq. (\ref{MEm}) the term ${\bf B}\times
\mathbf{m}_r$, where ${\bf B}= B\hat{\mathbf{z}}$. A
straightforward evaluation of $\langle \langle {\boldsymbol
\Omega} (t_1)\, {\boldsymbol \Omega} (t_2)\rangle \rangle$ by
repeating the steps that have led from Eq. (\ref{K2def}) to Eq.
(\ref{K2ans}) is insufficient; magnetic field effects appear only
in the fourth order, as a correction to $K_4$ of order $\sim\bigl(
B b_{\text{hf}}\bigr)^2$. Rather, a systematic expansion of $P(t)$
in powers of $b_{\text{hf}}$ can be achieved after performing the
rotating-frame transformation,
$\mathbf{m}_r(t)=\exp(t\,{\boldsymbol \Omega}_B) \mathbf{n}_r(t)$.
Then $\mathbf{n}_r(t)$ satisfies Eq.~(\ref{MEm}) with
time-dependent local frequencies, $\tilde{\bf b}_r(t)=
\exp(-t\,{\boldsymbol \Omega}_B){\bf b}_r \exp(t\,{\boldsymbol
\Omega}_B)$, and initial condition, $\mathbf{n}_r(0)=\delta_{r0}
\mathbf{m}(0)$. Also, as $\exp(t\,{\boldsymbol \Omega}_B)$ does
not change the $z$-component of the vector on which it acts,
$P(t)$ is expressed through $\mathbf{n}_r(t)$ exactly in the way
it was in terms of $\mathbf{m}_r(t)$. Averaging over the
distribution of local hyperfine fields now gives $\langle \langle
\hat{\tilde{ b}} (t_1)\, \hat{\tilde{ b}} (t_2)\rangle
\rangle_{zz}=-2 \,b_{\text{hf}}^2\,
\cos\bigl(B[t_1-t_2]\bigr)\langle\delta_{r(t_1),
r(t_2)}\bigr\rangle_{\text{rw}}$, which leads to Eq. (\ref{mfsc})
after applying Eq. (\ref{viaG}) and large-$(\nu t)$ expansion of
the resulting Bessel function.
\begin{figure}[h]
\vspace{-0.3cm}
\centerline{\includegraphics[width=80mm,angle=0,clip]{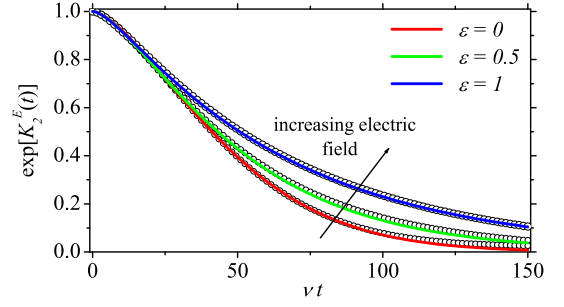}}
\vspace{-0.4cm} \caption{(Color online) $\exp[K_2^E(t)]$ is
plotted against $(\nu t)$ from Eq. (\ref{efsc}) with $\eta=0.05$,
for $\varepsilon=0$ (red), $\varepsilon=0.5$ (green), and
$\varepsilon=1$ (blue). Open black circles are simulated points
for $P(t)$ with corresponding values of parameters.}
\label{plotapp}
\end{figure}

In the presence of electric field along $\hat{\mathbf{x}}$, the
hopping rates forward and backward to $\hat{\mathbf{x}}$, $\nu_f$
and $\nu_b$, are different. In the paper we utilize $\nu_f=\nu/2$,
and $\nu_b=(\nu/2)e^{-\varepsilon}$. The diffusion propagator is
given by $q_r(t)=(\nu_f/\nu_b)^{\frac r 2} e^{-(\nu_f +\nu_b)t}
I_r\bigl(2\sqrt{\nu_f\nu_b}\,t\bigr)$.
This propagator leads to the second cumulant function,
\begin{widetext}
\begin{equation}\label{efsc}
K_2^E(t)= -2b_{\text{hf}}^2\!\int\limits_0^t\!\!dt_1\!\!
\int\limits_0^{t_1}\!\!dt_2 e^{-(\nu_f +\nu_b)[t_1-t_2]}
I_0\bigl(2\sqrt{\nu_f\nu_b}[t_1-t_2]\bigr)\simeq -\frac
{b_{\text{hf}}^2}{(\nu_f\nu_b)^{1/4}} \int\limits_0^t\!\!dt_1\!\!
\int\limits_0^{t_1}\!\!dt_2 \frac{e^{
-(\sqrt{\nu_f}-\sqrt{\nu_b})^2[t_1-t_2]}} {\sqrt{\pi(t_1-t_2)}}.
\end{equation}
The $E$-dependence of $K_2^E(t)$ is non-trivial because the
prefactor in Eq.~(\ref{efsc}) grows as $\exp(\varepsilon/4)$,
while the integral is suppressed with $E$. To clarify this
dependence and to show that, for almost all $t$, the absolute value of the cumulant
decreases with increasing $E$, we plot $\exp[K_2^E(t)]$ with
$\eta=0.05$ in Fig.~\ref{plotapp}, for three different values of
$E$.

To find the cumulant function $K_2^{>}(t)$ for the case of a
random walk over the semi-infinite chain, $r=1,2,...$, with the
reflecting boundary at $r=1$, we exploit its Greens function,
$G^{>}(r, r',t)=e^{-\nu t} \bigr[I_{|r-r'|}(\nu t)+I_{r+r'-1} (\nu
t)\bigl]$. Plugging this Greens function in Eq. (\ref{viaG}) and
using the large-$\nu t$ expansion of the modified Bessel function
yields
\begin{equation}\label{siomom}
\langle \langle {\boldsymbol \Omega} (t_1)\, {\boldsymbol \Omega}
(t_2)\rangle \rangle \simeq- \hat{\mathbf{1}}\cdot
\frac{2\,b_{\text{hf}}^2} {\sqrt{2\pi\nu}}\left( \frac
1{\sqrt{t_1-t_2}}+\frac1{\sqrt{t_1+t_2}}\right).
\end{equation}
By further integration we find the cumulant function Eq. (\ref{Kgcum}).

The regular lattice in $d=2$ can be described by the
radius-vector, $\mathbf{r}= (x, y)$, with $x, y=0,\pm1,\pm2,...$,
and $\hat{\mathbf{x}}$, $\hat{\mathbf{y}}$ along the lattice
sides. The Greens function of the random walk over this lattice is
$G^{(2)}(\mathbf{r},\mathbf{r}',t)=e^{-\nu t} I_{|x-x'|}\bigl(\nu
t/2\bigr)I_{|y-y'|}\bigl(\nu t/2\bigr)$. Further extension of Eqs.
(\ref{K2def})-(\ref{K2ans}) to $d=2$ is straightforward, leading
to
\begin{equation}\label{K22}
K_2^{(2)}(t)=-2\,b_{\text{hf}}^2 \int\limits_0^t\!dt_1\!
\int\limits_0^{t_1}\!dt_2\,e^{-\nu(t_1-t_2)}I_0^2\bigl(\nu(t_1-t_2)/2\bigr)
=- 2\,\eta^2\! \int\limits_0^{\nu t}\!dz(\nu
t-z)e^{-z}I_0^2\bigl(z/2\bigr).
\end{equation}
Combining numerical integration and asymptotic expansion of
$I_0(z)$ for large $z$, we find the large-$z$ expansion,
\begin{equation}\label{largez}
\int\limits_0^z\!dz'(z-z')e^{-z'}I_0^2\bigl(z'/2\bigr) = \frac
1\pi z\ln(\gamma z) -\frac1{2\pi}\ln(z) +\zeta +\mathcal{O}(1/z),
\qquad \gamma\approx 5.243,\quad \zeta\approx-0.264,
\end{equation}
yielding
\begin{equation}\label{K22fin}
K_2^{(2)}(t)\simeq -2\eta^2(\nu t)\ln(\gamma\nu t)/\pi.
\end{equation}

\end{widetext}


\begin{references}
\bibitem{footnote} Spin relaxation of a transiently diffusing carrier is
calculated analytically in Appendix~\ref{AppA}, by neglecting the
self-intersections of random walk trajectories.
\end{references}
\end{document}